\documentclass[conference]{IEEEtran}
\IEEEoverridecommandlockouts
\usepackage{cite}
\usepackage{amsmath,amssymb,amsfonts}
\usepackage{algorithmic}
\usepackage{graphicx}
\usepackage{textcomp}
\usepackage{xcolor}
\def\BibTeX{{\rm B\kern-.05em{\sc i\kern-.025em b}\kern-.08em
    T\kern-.1667em\lower.7ex\hbox{E}\kern-.125emX}}
\begin{document}

\title{Practical Radar Sensing Using Two Stage Neural Network for Denoising OTFS Signals
}

\author{
    \IEEEauthorblockN{%
        Ashok S Kumar, %
        $\, $%
        Sheetal Kalyani}
    \IEEEauthorblockA{\textit{Department of Electrical Engineering,} \\
        \textit{Indian Institute of Technology Madras,}%\\
        \textit{
            Chennai - 600036, %INDIA. \\
            India.
        }
        \\
        e-mail: \{ee22d023@smail, skalyani@ee\}.iitm.ac.in}
}

\maketitle

\begin{abstract}
Our objective is to derive the range and velocity of multiple targets from the delay-Doppler domain for radar sensing using orthogonal time frequency space (OTFS) signaling. Noise contamination affects the performance of OTFS signals in real-world environments, making radar sensing challenging. This work introduces a two-stage approach to tackle this issue. In the first stage, we use a generative adversarial network to denoise the corrupted OTFS samples, significantly improving the data quality. Following this, the denoised signals are passed to a convolutional neural network model to predict the values of the velocities and ranges of multiple targets. The proposed two-stage approach can predict the range and velocity of multiple targets, even in very low signal-to-noise ratio scenarios, with high accuracy and outperforms existing methods.
\end{abstract}

\begin{IEEEkeywords}
OTFS, delay-Doppler domain, convolutional neural network, generative adversarial network.
\end{IEEEkeywords}

\section{Introduction}
\label{sec:intro}

Orthogonal time frequency space (OTFS) signaling is emerging as a promising signal waveform for integrated sensing and communication systems \cite{hadani2017orthogonal}, \cite{wei2021orthogonal}. The OTFS signal modulates the data in the delay-Doppler (DD) domain. The target's range and velocity characteristics, derived from the DD domain, are the essential parameters to be calculated in radar signal processing. A 2D correlation-based approach to evaluate the Doppler and delay indices for radar sensing is studied in \cite{zhang2023radar}. The advantages of using OTFS waveform for velocity and range estimates in radar sensing applications have been investigated in \cite{muppaneni2023channel,9109735,shang2023target,yan2023deep,zacharia2023target,tan2024dnn}. A single target scenario is considered in \cite{muppaneni2023channel}, in which the root mean square error (RMSE) performance of the range and velocity as a function of signal-to-noise ratio (SNR) up to -15 dB is analyzed for radar target estimation. The work in \cite{9109735} mentions estimating the range and velocity RMSE as a function of radar SNR in a multipath scenario. The work reported in \cite{zacharia2023target}   exploited the application of three distinct sparse algorithms in estimating the range and velocity of the multiple targets using OTFS signaling.

In this work, we propose to improve the range and velocity RMSE of multiple targets by using a two-stage noise reduction method for OTFS signals. A generative adversarial network (GAN) is proposed for the first stage to denoise the corrupted OTFS samples.  The GAN-based denoising models have several advantages over traditional denoising methods \cite{singh2020new,an2022auto,wang2022ecg}. The advantages include better performance, enhanced generalization capabilities, the ability to exploit complex patterns in the data, and automation of the entire process. Next, a convolutional neural network (CNN) model is used to predict the
values of the delay and Doppler indices of multiple targets. These predicted values are used to calculate the RMSE for both velocity and range. 

In summary, this work presents a two-stage neural network model comprised of GAN and CNN to estimate the range and velocity for radar target detection using OTFS signaling. In real-world radar applications, the severity of noise in the signal is unpredictable, making radar target detection difficult. The existing literature falls short of providing a system dealing with radar target estimation in extremely noisy conditions. Simulation results show that our system can even operate in extremely noisy conditions ranging from -20 to 0 dB SNR, thus expanding the SNR range for radar target detection. Moreover, our proposed approach achieves better performance in terms of both velocity and range RMSE when compared to existing methods such as \cite{tan2024dnn,shang2023target,liu2024super,zhang2024target}.

\section{System Model}
\label{sec:system}

For each OTFS frame, $N$ and $M$ represent the number of time slots and the number of sub-carriers, respectively. The symbol duration is represented by $T$, and subcarrier frequency spacing is denoted by $\Delta f_{s}$. For a given $\Delta f_{s}$, the total bandwidth is $B=M \Delta f_{s}$
and the duration of one OTFS frame is given by $N T$. The information bits are mapped to a symbol set in the DD domain. The symbol set corresponding to $l^{th}$ delay and $k^{th}$ doppler bins is 
$ A_{\mathrm{DD}}[k,l]$, for $ k=0, \ldots, N-1$ and $l=0, \ldots, M-1 $. The DD domain symbols are  mapped to time-frequency (TF) domain symbols
using  inverse symplectic finite Fourier transform (ISFFT),
\begin{equation}
   A_{\mathrm{TF}}[n, m]=\frac{1}{\sqrt{N M}} \sum_{k=0}^{N-1} \sum_{l=0}^{M-1} A_{\mathrm{DD}}[k, l] e^{j 2 \pi\left(\frac{n k}{N}-\frac{m l}{M}\right)},
\end{equation}
where $ n=0, \ldots, N-1$ and $ m=0, \ldots, M-1 $. The grid $L$ in the TF domain is  a sampling of the frequency and time axis at intervals $\Delta f_{s}$ and $T$, respectively, where 
    $L= \{{(nT, m \Delta f_{s})}, n = 0, \ldots, N - 1; m = 0, \ldots, M-1\}$. The TF symbols are translated to the continuous time domain transmit
signal, $x(t)$ by using the Heisenberg transform,
\begin{equation}
\label{eqn:3}
x(t)=\sum_{n=0}^{N-1} \sum_{m=0}^{M-1} A_{\mathrm{TF}}[n, m] g_{t x}(t-n T) e^{j 2 \pi m \Delta f_{s}(t-n T)},
\end{equation}
where $g_{t x}$ is a rectangular pulse-shaping waveform.
The time domain signal $x(t)$ is passed through the linear time-varying channel, which has $P$ targets in the DD domain. The $p^{th}$ target has complex gain $h_{p}$, delay $\tau_{p}$, and  Doppler shift $\nu_{p}$ \cite{raviteja2019orthogonal}. The complex gain $h_{p}$ is assumed to follow a complex normal distribution, $\mathcal{CN}(0, 1/P)$ \cite{naikoti2021dnn}.
The complex base-band channel response, $h(\tau, \nu)$ in the DD domain is given by,
\begin{equation}
\label{eqn:4}
    h(\tau, \nu)=\sum_{p=0}^{P-1} h_{p} \delta \left(\tau-\tau_{p}\right) \delta \left(\nu-\nu_{p}\right).
\end{equation}
For integer delays and Doppler,
$\tau_{p}=\frac{l_{p}}{M \Delta f_{s}}$ and $\nu_{p}=\frac{k_{p}}{N T},
$ where $l_{p}$ and $k_{p}$  denote the corresponding delay and Doppler indices of the $p^{th}$ target. The received signal $r(t)$ is given by,
% \textcolor{blue}{\textbf{We assume a uniform gain ($h_{p}=1$)  for each multipath component (or each target), which helps isolate and study the effects of delay and Doppler shifts on the system's performance.}} The received signal $r(t)$ is given by,
\begin{equation}
\label{eqn:5}
r(t)=\iint h(\tau, \nu) e^{j 2 \pi \nu(t-\tau)} x(t-\tau) d \tau d \nu+w(t),
\end{equation}
where $w(t)$ denotes the additive white complex Gaussian noise (AWGN) process with one side power spectral density (PSD), $N_{0}$ \cite{zhang2023radar}. The received
signal $r(t)$ is converted back to the TF domain using Wigner
transform,
\begin{equation}
B_{\mathrm{TF}}[n, m]=\int_{-\infty}^{\infty} r(t) g_{r x}^{*}(t-n T) e^{-j 2 \pi m \Delta f_{s}(t-n T)} d t,
\end{equation}
where $g_{r x}(t)$ is the rectangular pulse-shaping filter at the receiver. 
The TF domain signals $B_{\mathrm{TF}}[n, m]$ are then converted to DD domain
symbols $B_{\mathrm{DD}}[k, l]$ using symplectic finite Fourier transform (SFFT), which is given by
\begin{equation}
  B_{\mathrm{DD}}[k, l]=\frac{1}{\sqrt{N M}} \sum_{n=0}^{N-1} \sum_{m=0}^{M-1} B_{\mathrm{TF}}[n, m] e^{-j 2 \pi\left(\frac{n k}{N}-\frac{m l}{M}\right)}.
\end{equation}
In view of the fact that $B_{\mathrm{DD}}$ contains information symbols, we are not able to identify the target areas of interest directly. Instead, a 2D correlation-based approach has been used between $B_{\mathrm{DD}}$ and $A_{\mathrm{DD}}$ to obtain the delay and Doppler index \cite{zhang2023radar}. 
The matrix $V$ contains information about the correlation between the transmitted and received signals at different delay and Doppler indices. 
The accumulated correlation coefficient under different delay
and Doppler indices is given by,
\begin{equation}
   \begin{gathered}
   \label{eqn:8}
V[k, l]=\sum_{n=0}^{N-1} \sum_{m=0}^{M-1} B_{\mathrm{DD}}^{*}[n, m] A_{\mathrm{DD}}\left[[n-k]_{N},[m-l]_{M}\right] \\
\times \gamma[n-k, m-l] e^{j 2 \pi \frac{(m-l) k}{N M}},
\end{gathered}
\end{equation}
where $k \in[0, N-1]$ and $l \in[0, M-1]$, and $\gamma[k, l]$ is a phase offset given by 
\begin{equation} 
\gamma[k, l]= \begin{cases}1, & l \geq 0, \\ e^{-j 2 \pi \frac{k}{N}}, & l<0 .\end{cases}
\end{equation}

\section{Proposed Approach}
\label{sec:Prop}
We propose a two-stage neural network comprising a GAN and a CNN. We first describe the datasets we use to train the proposed deep learning model.
\begin{itemize}
\item \textbf{2D corrupted dataset:} The 2D corrupted dataset contains the DD matrices for radar sensing after performing the 2D correlation between corrupted/noisy signal and transmitted signal in the DD domain. The 2D corrupted dataset is obtained using the equation (\ref{eqn:8}). The transmitted OTFS signal $x(t)$ in equation (\ref{eqn:3}) is used to probe the environment and detect the objects or targets that reflect the signal to the receiver. The corrupted signal $r(t)$ is obtained from equation (\ref{eqn:5}) by adding complex AWGN noise, where SNR ranges from -20 to 0 dB.

\item \textbf{2D clean dataset:} The 2D clean dataset contains the DD matrices for radar sensing after performing the 2D correlation between low noise signal and transmitted signal in the DD domain. This is obtained using the equation (\ref{eqn:8}). Here, the low noise signal $r(t)$ is obtained by using the equation (\ref{eqn:5}) by adding AWGN noise of 20 dB SNR. Typically, deep learning applications operate under the assumption of a completely clean dataset with no noise.  However, in practical scenarios, such datasets are rarely available. Hence, we use low-noise signals in our work. We have created clean datasets with 20 dB SNR value. These datasets are used as the input to the GAN only during the training phase.

\item \textbf {Label:} The labels used for training the CNN are the targets' true delay and Doppler indices in the DD domain. The delay index ($l_{p}$) corresponds to the target's delay in the DD domain, which is related to the target's distance and the Doppler index ($k_{p}$) corresponds to the target's Doppler shift, which is related to the target's velocity.

\end{itemize}

\subsection{First Stage: Denoising OTFS Signals Using GAN}
\label{subsec:GAN}
GANs use two neural networks, a generator network $G$ and a discriminator network $D$, competing against each other to create the desired result. The inputs 
to the discriminator and generator are actual data $u$ and random variable $w$, respectively. The discriminator gives a value $D(u)$, suggesting the possibility that $u$ is a real sample. The main purpose of the discriminator is to maximize the probability of labeling the real samples as real and the generated fake samples $G(w)$ as fake. The objective of the generator is to produce fake samples $G(w)$, which are as close as possible to that of real samples so that the discriminator fails to separate between fake and real samples. Hence, a  GAN can be defined as a minimax game in which $G$ wants to minimize the value function $\Tilde{V}(D, G)$, while $D$ wants to maximize it \cite{goodfellow2020generative}.
\begin{equation}
\min _G \max _D \Tilde{V}(D, G)=\mathbb{E}[\log D(u)]+\mathbb{E}[\log (1-D(G(w)))]  
\end{equation}

Fig. 1. shows the block diagram of GAN for denoising OTFS signals. The generator network is trained to generate denoised signals from the 2D corrupted signals. The input to the discriminator network is a 2D clean dataset and generated signals from the generator. During training, the generator network attempts to minimize the difference between the generated and clean signals, while the discriminator network aims to maximize the difference between the generated and clean signals. The 2D corrupted dataset is normalized before feeding to the generator, ensuring consistent input scaling, ultimately improving model performance. The network starts with an input layer of dimension (28 × 28) with a single channel. This layer is followed by two convolutional layers with $64$ and $128$ filters, each with size (3,3). The padding is set to be one in both cases. Both these layers are followed by a batch normalization and a ReLU activation function. The final layer is again a convolutional layer that reduces the number of feature maps back to one, producing a single-channel output image. The final layer uses a tanh activation function to produce an output image of the same shape as the input with pixel values in the range [-1, 1]. It represents the generated signals, which are then fed to the discriminator. The discriminator network starts with an input layer ($28\times28$) followed by the convolution layer, which has $64$ filters of size $(4,4)$ and a stride of $2$. The padding is set to one. The output of the convolution layer is then fed through a leaky ReLU activation function with a negative slope of 0.2. The same convolution layers are repeated with 128 filters and 256 filters of size $(4,4)$ and $(3,3)$, respectively (stride=2). Both these layers are followed by a batch normalization and leaky ReLU activation function with a negative slope of 0.2. The output of these is flattened and passed through a dense layer with a sigmoid activation function. The model is compiled with Adam optimizer (learning rate=0.0002) and binary cross-entropy loss.

\begin{figure}[hb]
\includegraphics[width=3.3 in]{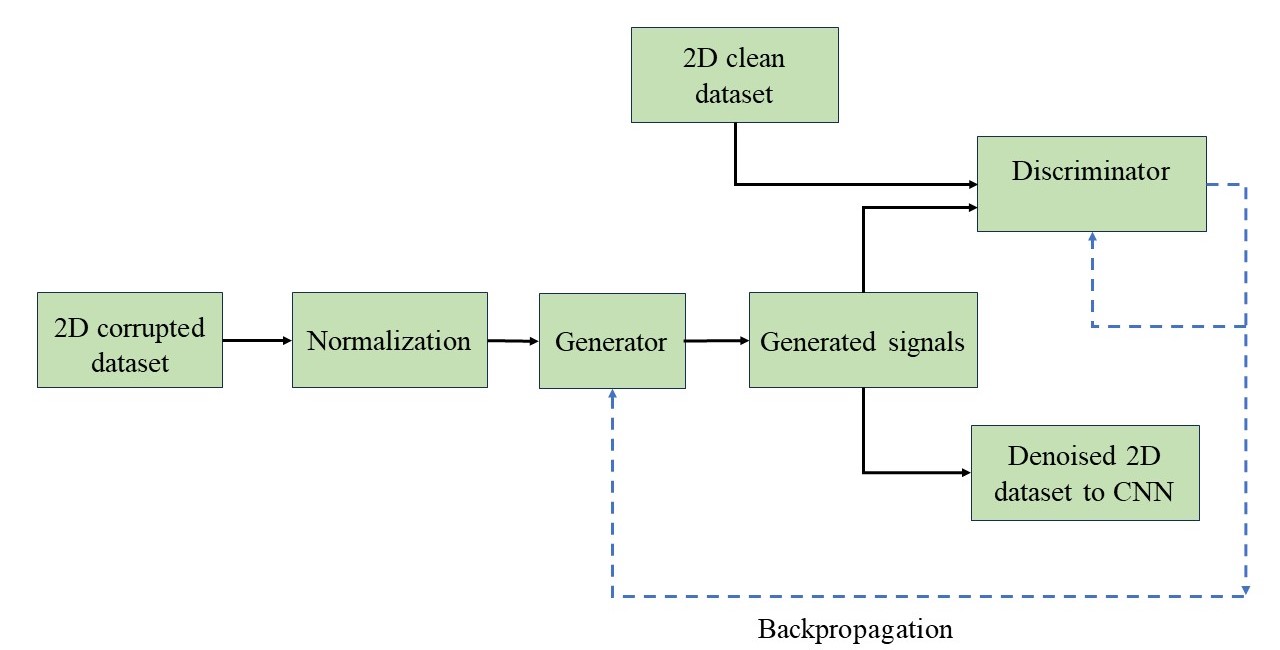}
\label{fig3}
\centering
   \caption{Block diagram of GAN for denoising}
\end{figure}

\begin{figure*}[!ht]
\includegraphics[width=6.4in,angle= 0]{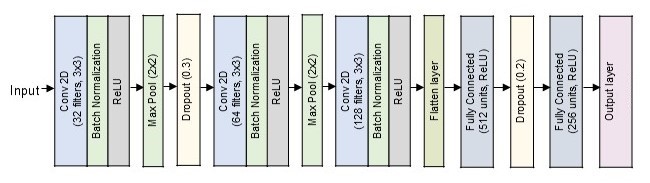}
\label{fig22}
\centering
    \caption{Proposed CNN architecture for predicting the range
and velocity of the target}

\end{figure*}

\subsection{Second Stage: CNN Architecture for Predicting the Range and Velocity of the Target}
\label{subsec:CNN}
The denoised signals from the GAN (generator) are fed to the proposed  CNN to predict the range and velocity of the targets. Fig. 2. shows the proposed CNN architecture. The network starts with an input layer of dimension ($28 \times 28$). This layer is succeeded by a convolution 2D layer (conv2D) with 32 filters of size ($3 \times 3$).
The padding is set to be the same. The batch normalization layer is inserted between the layers which is then followed by the ReLU layer. To reduce the spatial dimension, a maxpooling layer with a pool size of ($2 \times 2$) is used, followed by a dropout layer (0.3 dropout rate) to prevent overfitting. The CNN then continues with a series of 2 similar layers alternating between convolutional layers, batch normalization layers, and ReLU layers. The filter size gradually increases from 32, then 64 and finally reaching 128 in the subsequent layers. The flattened output is then given to two fully connected dense layers with $512$ and $256$ units, respectively. The model is compiled with the Adam optimizer. The final fully connected layer outputs the values corresponding to the predicted delay and Doppler indices of the multiple targets. From this, we can estimate the velocities and ranges of the targets. The range $R_{p}$ and velocity $V_{p}$ can be derived from 
$\tau_{p}=\frac{2 R_{p}}{c}$ and  $\nu_{p}=\frac{2 f_{c} V_{p}}{c}.
$

\section{Simulation Results}

\label{sec:result}
A DD domain grid with $M = N = 28$ and $\Delta f_{s}$ = 150 kHz is considered. The carrier frequency, $f_{c}$ is taken as 60 GHz. The velocity and range resolution can be
calculated by $ V_{res} =\frac{Bc}{2MNf_{c}}$ and $R_{res} =\frac{c}{2B}$, where $c$ is the speed of light. The maximum velocity and range are given by $V_{max}=\frac{c\Delta f_{s}}{2f_{c}}$ and $R_{max}=\frac{cT}{2}$. The range RMSE is given by,
\begin{equation}
    R_{RMSE}=R_{\text{res}} \cdot \sqrt{\frac{1}{N_{\text{S}} \cdot P} \sum_{j=1}^{N_{\text{S}}} \sum_{i=1}^{P} \left( \tau_{\text{tr}}^{(i,j)} - \tau_{\text{pr}}^{(i,j)} \right)^2}.
\end{equation}
Similarly, the velocity RMSE is given by,

\begin{equation}
    V_{RMSE}=V_{\text{res}} \cdot \sqrt{\frac{1}{N_{\text{S}} \cdot P} \sum_{j=1}^{N_{\text{S}}} \sum_{i=1}^{P} \left( \nu_{\text{tr}}^{(i,j)} - \nu_{\text{pr}}^{(i,j)} \right)^2}.
\end{equation}
Here $N_{S}$ is the number of samples. The true delay and predicted delay indices are denoted by $\tau_{\text{tr}}^{(i,j)} \quad \text{(for the $i^{th}$ target in the $j^{th}$ sample)}$ and 
$\tau_{\text{pr}}^{(i,j)}$. Also, $
\nu_{\text{true}}^{(i,j)}$ and
$\nu_{\text{true}}^{(i,j)}$ denotes the true Doppler and predicted Doppler indices. The dataset comprising 50000 complex OTFS samples, each with a length of $MN$, is used to obtain the low noise, corrupted, and transmitted signals. From this, we obtain the corresponding 2D clean dataset and 2D corrupted dataset. Fig. 3. shows the DD matrices after performing the 2D correlation between the low noise signal and transmitted signal in the DD domain. In this example, two targets are considered. The delay and Doppler indices of the first target are (2,10). The delay and Doppler indices of the second target are (7,17). Fig. 4. shows the DD matrices after performing the 2D correlation between the corrupted signal and the transmitted signal in the DD domain. We cannot identify the location of the target exactly from the DD domain matrix.

\begin{figure}[ht]

\includegraphics[width=2.8 in]{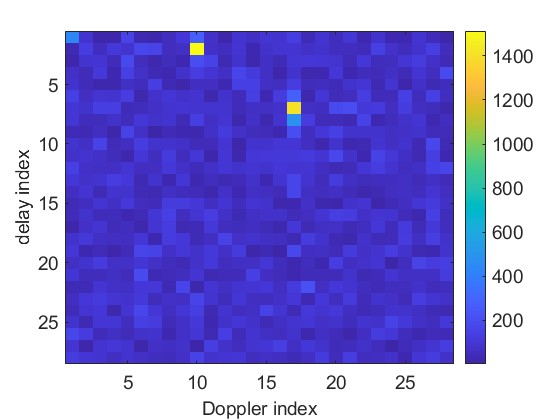}
\label{fig1}
\centering

\caption{The DD matrices obtained by performing the 2D correlation between low noise signal and transmitted signal in the DD domain. This 2D clean data is given as input to the discriminator.} 
% The below plot corresponds to the 3D plot by taking the magnitude of $V$ along the $z-$axis}
\end{figure}

%where $P=1$ targets are considered and the delay and Doppler indices are of integer values

\begin{figure}[ht]
\includegraphics[width=2.8 in]{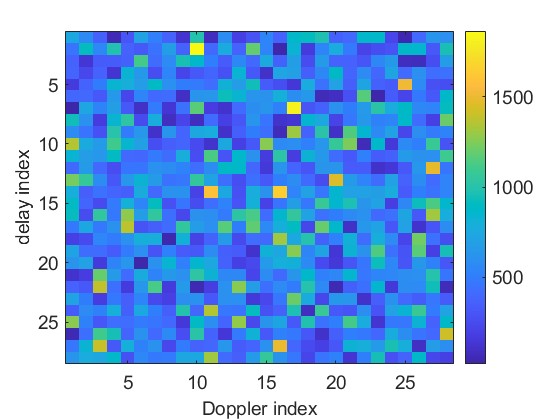}
\label{fig33}
\centering
\centering
    \caption{The DD matrices obtained by performing the 2D correlation between corrupted signal and transmitted signal in the DD domain. This 2D corrupted data is given as input to the generator.}
\end{figure}

\subsection{Performance Analysis and Comparison With the State-of-the-Art Methods}
\label{subsec:validation}
For testing the system, 10000 2D clean datasets and 2D corrupted datasets are given to GAN. The denoised generated samples are given to the proposed CNN to predict the delay and Doppler index of the targets. Our system can support a maximum distance of 1000 m and a maximum velocity of 375 m/s. When tested with two targets under -20 dB SNR, we obtained a range RMSE and a velocity RMSE of 22.49 m and 14.7 m/s, respectively. Additionally, when tested with 2D corrupted datasets at an SNR of -15 dB with two targets, we obtain a range RMSE and a velocity RMSE of 11.68 m and 8.43 m/s. The work \cite{shang2023target} addresses modified versions of maximum likelihood estimation algorithms for multiple target detection using OTFS signaling. The paper \cite{shang2023target} considers three targets at -15 dB SNR with maximum range and velocity of 240 m and 254.237 m/s, respectively. For these targets, the range RMSE (at an SNR of up to -15 dB) is 12 m and the velocity RMSE is 10 m/s. When we tested with three target scenarios at -15 dB SNR, for a maximum distance of 1000 m and a maximum velocity of 375 m/s, we obtained a range RMSE and a velocity RMSE of 23.92 m and 10.71 m/s. This indicates that our proposed system maintains low RMSE values while supporting higher ranges and velocities. The work in \cite{liu2024super} presents a super-resolution approach for joint delay-Doppler estimation in automotive radar using OTFS modulation. The study considers four targets with maximum range and velocity of 960 m and 304 m/s. The paper reports range and velocity RMSE values of 108 m and 30 m/s at -20 dB SNR. At this SNR, with four targets, we obtained a range RMSE and a velocity RMSE of 61.06 m and 24.50 m/s which is significantly better. In \cite{zhang2024target}, a parameter estimation method for OTFS-integrated radar and communications systems, utilizing sparse reconstruction preprocessing is proposed. This method achieves a range RMSE of 10 m for three targets at a maximum range of 240 meters, with an SNR of -20 dB. In contrast, our proposed approach, tested with three targets at the same SNR, achieves a range RMSE of 28.92 m, but for a significantly larger maximum range of 1000 meters. The work in \cite{tan2024dnn} proposed a DNN-based radar target detection method with OTFS, designed to handle multiple targets across varying SNR levels, ranging from 15 dB to -5 dB. Due to our dual-stage technique, we can work even with SNR as low as -20 dB.  While previous works exhibit significant RMSE values, our method achieves notably lower RMSE values even for higher range and velocity, showcasing its robustness in handling noisy datasets. To summarize, accurate radar target detection in highly challenging, low SNR environments is now possible due to our proposed approach.

 % \textbf{In \cite{muppaneni2023channel}, a single target scenario up to -15 dB shows a range RMSE of 0.5 m and a velocity RMSE of 3 m/s.}

\section*{Conclusions}
In this work, we have proposed a two-stage approach for denoising OTFS signals for radar sensing. The proposed two-stage approach can be used to derive the range and velocity of multiple targets even under very low SNR scenarios. The first step involves the denoising of noisy OTFS signals using GAN. Subsequently, the next stage involves a CNN model to predict
the values of the velocities and ranges of multiple targets. The proposed system has yielded promising results demonstrating its effectiveness in both the denoising and prediction of the range and velocity of multiple targets, even in very low SNR environments.

\bibliographystyle{IEEEtran}
\bibliography{ref}

\begin{thebibliography}{10}
\providecommand{\url}[1]{#1}
\csname url@samestyle\endcsname
\providecommand{\newblock}{\relax}
\providecommand{\bibinfo}[2]{#2}
\providecommand{\BIBentrySTDinterwordspacing}{\spaceskip=0pt\relax}
\providecommand{\BIBentryALTinterwordstretchfactor}{4}
\providecommand{\BIBentryALTinterwordspacing}{\spaceskip=\fontdimen2\font plus
\BIBentryALTinterwordstretchfactor\fontdimen3\font minus \fontdimen4\font\relax}
\providecommand{\BIBforeignlanguage}[2]{{%
\expandafter\ifx\csname l@#1\endcsname\relax
\typeout{** WARNING: IEEEtran.bst: No hyphenation pattern has been}%
\typeout{** loaded for the language `#1'. Using the pattern for}%
\typeout{** the default language instead.}%
\else
\language=\csname l@#1\endcsname
\fi
#2}}
\providecommand{\BIBdecl}{\relax}
\BIBdecl

\bibitem{hadani2017orthogonal}
R.~Hadani, S.~Rakib, M.~Tsatsanis, A.~Monk, A.~J. Goldsmith, A.~F. Molisch, and R.~Calderbank, ``{Orthogonal time frequency space modulation},'' in \emph{2017 IEEE Wireless Communications and Networking Conference (WCNC)}, 2017, pp. 1--6.

\bibitem{wei2021orthogonal}
Z.~Wei, W.~Yuan, S.~Li, J.~Yuan, G.~Bharatula, R.~Hadani, and L.~Hanzo, ``{Orthogonal time-frequency space modulation: A promising next-generation waveform},'' \emph{IEEE wireless communications}, vol.~28, no.~4, pp. 136--144, 2021.

\bibitem{zhang2023radar}
K.~Zhang, W.~Yuan, S.~Li, F.~Liu, F.~Gao, P.~Fan, and Y.~Cai, ``{Radar sensing via OTFS signaling: A delay doppler signal processing perspective},'' in \emph{ICC 2023-IEEE International Conference on Communications}, 2023, pp. 6429--6434.

\bibitem{muppaneni2023channel}
S.~P. Muppaneni, S.~R. Mattu, and A.~Chockalingam, ``{Channel and radar parameter estimation with fractional delay-Doppler using OTFS},'' \emph{IEEE Communications Letters}, vol.~27, no.~5, pp. 1392--1396, 2023.

\bibitem{9109735}
L.~Gaudio, M.~Kobayashi, G.~Caire, and G.~Colavolpe, ``{On the Effectiveness of OTFS for Joint Radar Parameter Estimation and Communication},'' \emph{IEEE Transactions on Wireless Communications}, vol.~19, no.~9, pp. 5951--5965, 2020.

\bibitem{shang2023target}
X.~Shang, Z.~Zhang, and Y.~Xiao, ``{A target parameter estimation algorithm for integration of radar and communication based on orthogonal time frequency space},'' \emph{IET Radar, Sonar \& Navigation}, vol.~17, no.~7, pp. 1142--1151, 2023.

\bibitem{yan2023deep}
Z.~Yan, L.~Tan, X.~Zhang, K.~Zhang, and R.~Zhou, ``{Deep Learning-Assisted Target Classification Using OTFS Signaling},'' in \emph{2023 IEEE/CIC International Conference on Communications in China (ICCC Workshops)}, 2023, pp. 1--6.

\bibitem{zacharia2023target}
O.~Zacharia and M.~V. Devi, ``{Target parameter estimation for OTFS radar using sparse signal processing},'' \emph{Physical Communication}, vol.~58, p. 102040, 2023.

\bibitem{tan2024dnn}
L.~Tan, W.~Yuan, X.~Zhang, K.~Zhang, Z.~Li, and Y.~Li, ``{DNN-Based Radar Target Detection With OTFS},'' \emph{IEEE Transactions on Vehicular Technology}, pp. 1--6, 2024.

\bibitem{singh2020new}
P.~Singh and G.~Pradhan, ``{A new ECG denoising framework using generative adversarial network},'' \emph{IEEE/ACM Transactions on Computational Biology and Bioinformatics}, vol.~18, no.~2, pp. 759--764, 2020.

\bibitem{an2022auto}
Y.~An, H.~K. Lam, and S.~H. Ling, ``{Auto-Denoising for EEG Signals Using Generative Adversarial Network},'' \emph{Sensors}, vol.~22, no.~5, p. 1750, 2022.

\bibitem{wang2022ecg}
X.~Wang, B.~Chen, M.~Zeng, Y.~Wang, H.~Liu, R.~Liu, L.~Tian, and X.~Lu, ``{An ECG Signal Denoising Method Using Conditional Generative Adversarial Net},'' \emph{IEEE Journal of Biomedical and Health Informatics}, vol.~26, no.~7, pp. 2929--2940, 2022.

\bibitem{liu2024super}
S.~Liu, H.~Zhang, L.~Li, Z.~Gong, Y.~Huang, and J.~Yuan, ``{Super-resolution delay-Doppler estimation for OTFS-based automotive radar},'' \emph{Signal Processing}, vol. 224, p. 109596, 2024.

\bibitem{zhang2024target}
Z.~Zhang, X.~Shang, and Y.~Xiao, ``{Target parameter estimation for OTFS-integrated radar and communications based on sparse reconstruction preprocessing},'' \emph{Frontiers of Information Technology \& Electronic Engineering}, vol.~25, no.~5, pp. 742--754, 2024.

\bibitem{raviteja2019orthogonal}
P.~Raviteja, K.~T. Phan, Y.~Hong, and E.~Viterbo, ``{Orthogonal time frequency space (OTFS) modulation based radar system},'' in \emph{2019 IEEE Radar Conference (RadarConf)}, 2019, pp. 1--6.

\bibitem{naikoti2021dnn}
A.~Naikoti and A.~Chockalingam, ``{A DNN-based OTFS transceiver with delay-Doppler channel training and IQI compensation},'' in \emph{2021 IEEE 32nd Annual International Symposium on Personal, Indoor and Mobile Radio Communications (PIMRC)}, 2021, pp. 628--634.

\bibitem{goodfellow2020generative}
I.~Goodfellow, J.~Pouget-Abadie, M.~Mirza, B.~Xu, D.~Warde-Farley, S.~Ozair, A.~Courville, and Y.~Bengio, ``{Generative adversarial networks},'' \emph{Communications of the ACM}, vol.~63, no.~11, pp. 139--144, 2020.

\end{thebibliography}

\end{document}